\documentclass{PoS}

\pdfoutput=1 

\usepackage[round]{natbib}
\shortcites{mittal06}

\usepackage{journals}

\usepackage{url}

\newcommand{\sub}[1]{_{\rm#1}}

\newcommand{\hlinkbibcode}[1]{http://adsabs.harvard.edu/cgi-bin/bib_query?#1}

\title{Gravitational lensing}

\ShortTitle{Gravitational lensing}

\author{\speaker{Olaf Wucknitz}%
\thanks{Current address: Argelander-Institut f\"ur Astronomie, Universit\"at
  Bonn, Germany}
\\
        Joint Institute for VLBI in Europe, Dwingeloo, Netherlands \\
        E-mail: \email{wucknitz@astro.uni-bonn.de}}

\abstract{This is a short and biased review of gravitational lensing with 
  emphasis on the radio and especially VLBI aspects. We briefly explain the
  basic idea and give a short sketch of the discovery of the first lens before
  we more systematically discuss the general fields that can be studied with
  lensing. We intentionally omit the details the average lensing expert would
  like to see and instead try to give a very general overview addressed to the
  radio astronomer working in some other field.
  The lens B0218+357 is presented as an example to show many aspects of
  lensing in a case that already has led to interesting results but still has
  additional potential for the future.}

\FullConference{8th European VLBI Network Symposium\\
                 September 26-29, 2006\\
                 Toru\'n, Poland}

\begin{document}

\section{The idea}

\begin{figure}[htb]
\centering
\ifpdf%
\includegraphics[angle=-90,width=0.5\textwidth]{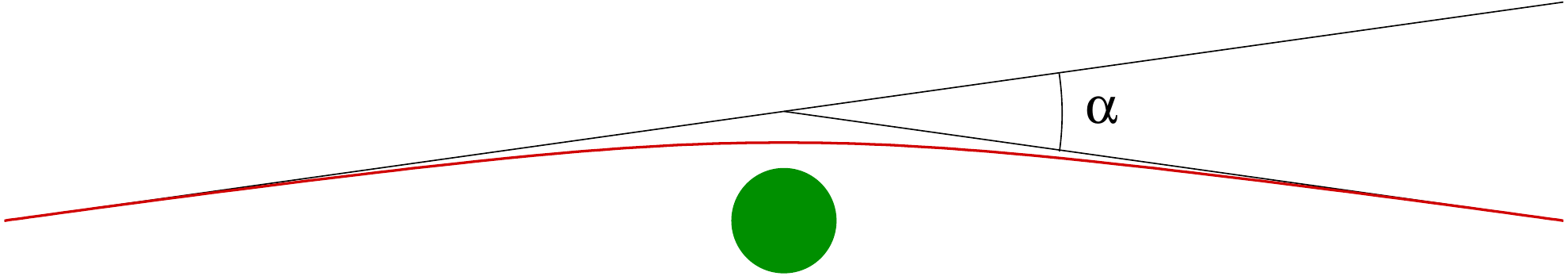}%
\else%
\includegraphics[width=0.5\textwidth]{ablenk}%
\fi%
\caption{Light deflection}
\label{fig:defl}
\end{figure}

The idea that gravity might influence the propagation of light is much older
than the EVN or even radio astronomy. Even \citet{newton04} mentioned the
possibility\footnote{``Do not bodies act upon light at a distance,
and by their action bend its rays; and is not this action strongest at the
least distance?''}, but without going into any details.
Henry Cavendish was the first to calculate deflections using the simplified
model that light consists of classical particles moving with the speed of
light, thus being weakly deflected in a gravitational field
\citep[see][]{will88}. Without 
knowing of this unpublished work, \citet{soldner1801} used a similar concept and
wrote the first article about gravitational lensing.
He found that the deflection angle (see Fig.~\ref{fig:defl}) between the
asymptotic light paths 
before and after an encounter with a mass $M$ at a distance $r$ is
$\alpha=2GM/(c^2r)$, where $G$ and $c$ are the gravitational constant and the
speed of light, respectively. Before developing the theory of general
relativity, \citet{einstein11} used another approach that does not rely
on the incorrect classical model of light but, instead, used the principle of
equivalence to find the same result as Soldner.
A few years later, our still best theory of gravitation was finished and
\citet{einstein15} could reconsider the deflection of light, finding that it
should be \emph{twice} as large as in non-relativistic theory:
\begin{equation}
 \alpha = \frac{4GM}{c^2 r}
\label{eq:defl}
\end{equation}
This result was confirmed (with limited accuracy) by \citet{eddington19}, who
determined the deflection of light caused by the Sun by measuring positions
of stars close to its limb during a solar eclipse. This
served as a first independent test of the new theory.

Almost all aspects of gravitational lensing theory can be derived from
Eq.~(\ref{eq:defl}), together with the astrophysical background that is
required to relate it to observable quantities. 
It is easy to see that in the presence of sufficiently large and compact
masses, situations are possible where light from a background source reaches
the observer along different light paths, which are then interpreted as
multiple images of this source. Typical lenses show two or four images, which
in very symmetrical systems merge to form rings around the lens, the
so-called Einstein rings.

\section{The first lensed double: 0957+561}

\begin{figure}[htb]
\centering
\raisebox{1.7ex}{\includegraphics[width=0.47\textwidth]{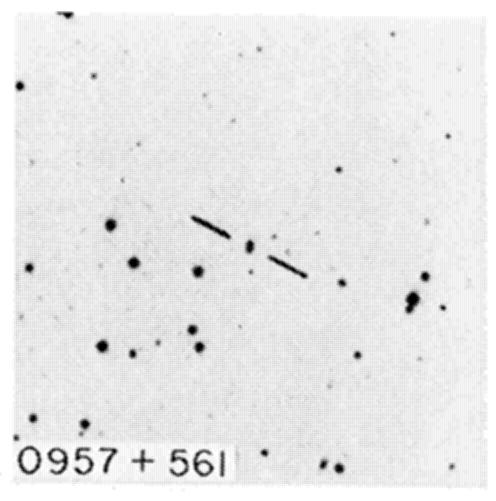}}%
\llap{\raisebox{0.29\textwidth}{\includegraphics[width=0.18\textwidth]{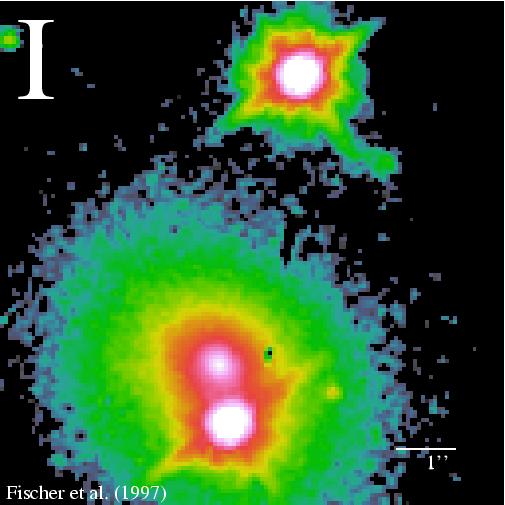}}\hspace*{0.6em}}%
\hfill%
\raisebox{0.8ex}{\includegraphics[width=0.47\textwidth]{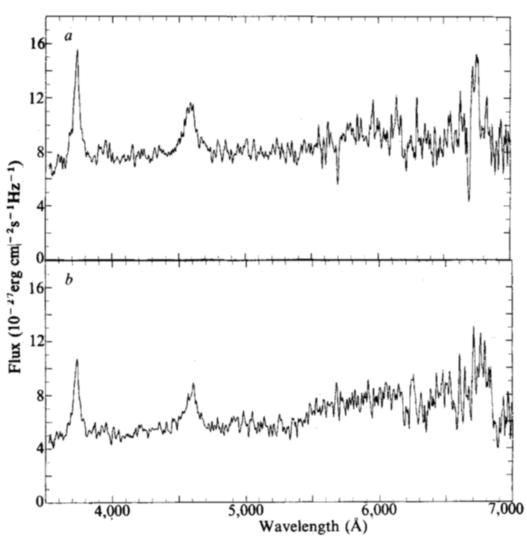}}
\caption{Left: Optical POSS image and identification of 0957+561
  \citep[from][]{porcas80}. Inset: HST I image \citep[from][]{castles}
showing the two images \emph{A} and \emph{B} and the lensing galaxy \emph{G}
close to \emph{B}. 
Right: Spectra of both images \citep[from][]{walsh79}}
\label{fig:0957}
\end{figure}

This radio source (without knowing about its lensed nature at that time) was
discovered in a 966-MHz survey made with the Jodrell Bank MkIA telescope.
One of the sources in that survey, 0958+56, was a blend of the spiral
galaxy NGC~3079 with the new source 0957+561. The latter had an optical
identification as two point-like images with a separation of about $6''$
\citep[and Fig.~\ref{fig:0957}, left]{porcas80}.

Optical spectra of the two images appeared like almost identical copies of a
QSO spectrum with a redshift of $z=1.4$. Since a chance alignment of so
similar objects with such a close separation is extremely unlikely, the
favoured interpretation was (and still is) that we see two gravitationally
lensed images of one and the same source \citep{walsh79}.
Many interesting (and amusing) details of the first discovery, including a
description of all coincidences that had to conspire to allow the success, are
described by \citet{walsh89}.

\begin{figure}[htb]
\centering
\includegraphics[height=0.27\textwidth]{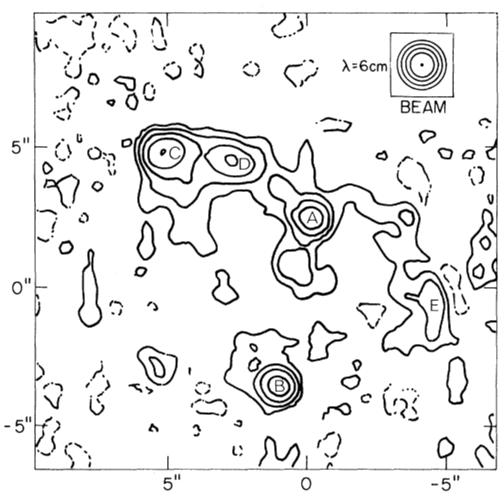}%
\hfill%
\includegraphics[height=0.28\textwidth]{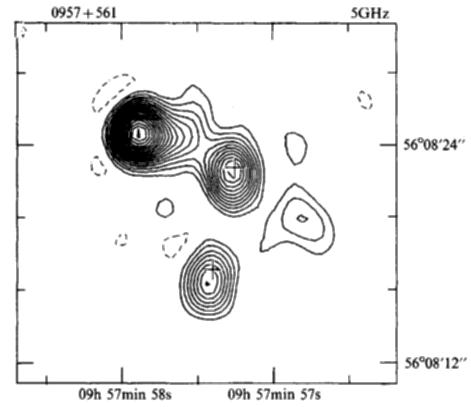}%
\hfill%
\includegraphics[height=0.27\textwidth]{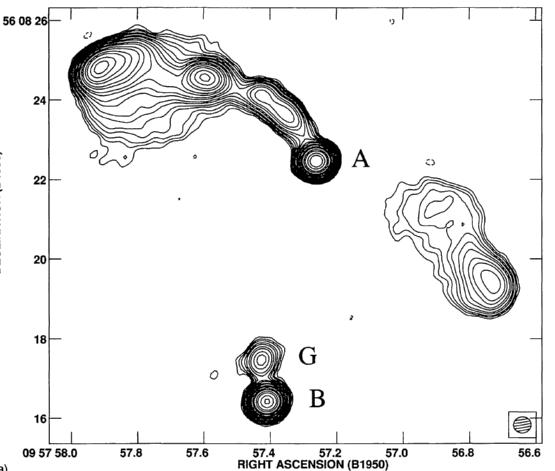}
\caption{Interferometric observations of 0957+561 at 6\,cm. Left: early VLA
  \citep{roberts79}. Centre: Cambridge 5\,km. Right: full VLA
  \citep{harvanek97}.}
\label{fig:0957 interf}
\end{figure}

\begin{figure}[htb]
\begin{minipage}{0.36\textwidth}
\ifpdf%
\includegraphics[angle=-90,width=\textwidth]{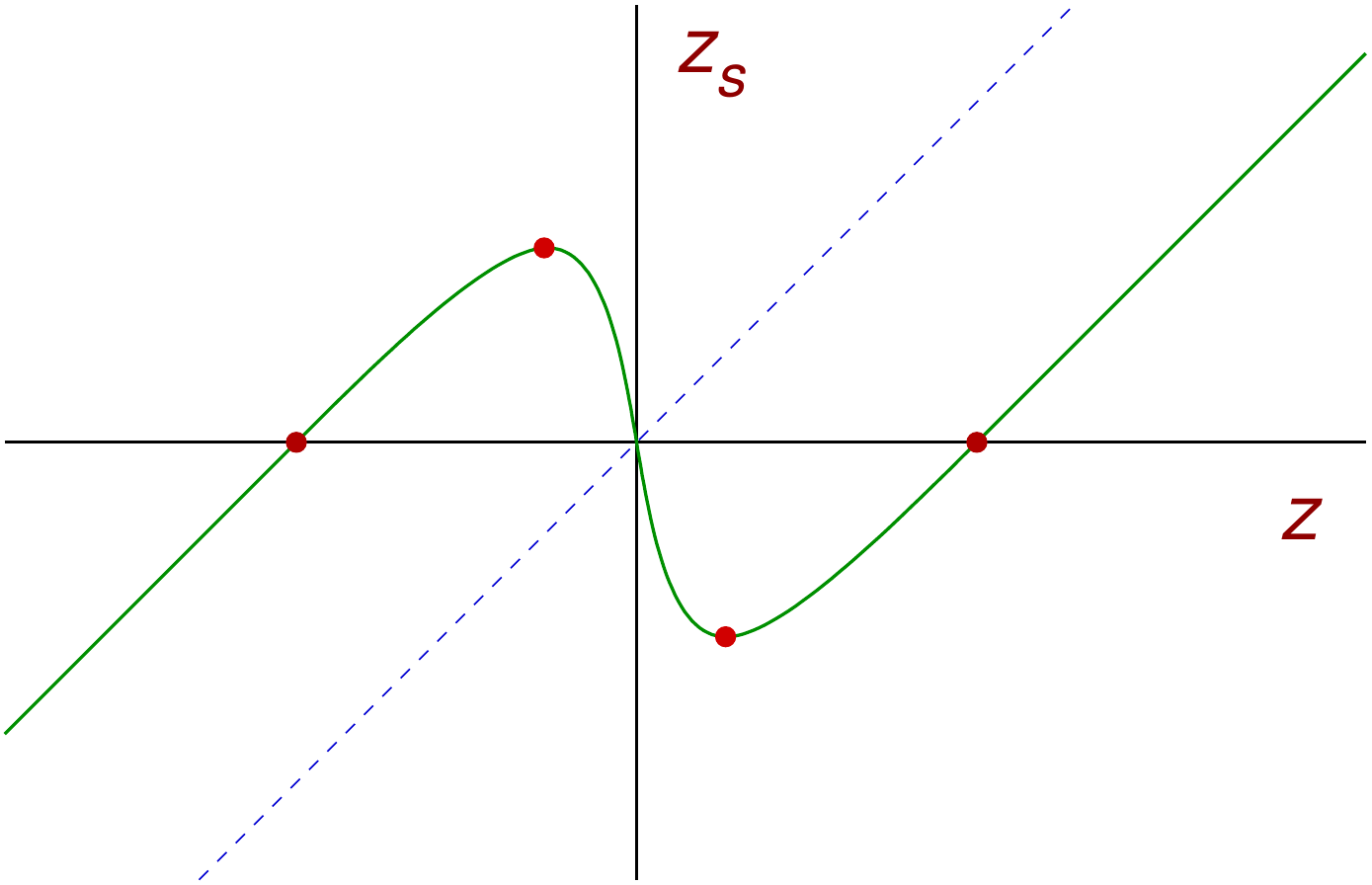}%
\else%
\includegraphics[width=\textwidth]{diag}%
\fi
\end{minipage}\hfill\begin{minipage}{0.61\textwidth}
\includegraphics[angle=180,width=\textwidth]{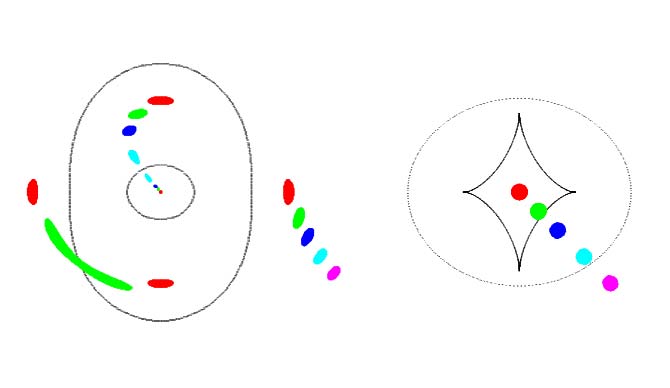}
\end{minipage}
\caption{Left: Illustration of a typical lens mapping in one dimension,
  showing the true source position $z\sub s$ as a function of the
apparent image position $z$. For source positions far away from the origin, we
see only one image. When the source crosses the extrema of the curve, two
additional images appear or disappear.
Centre and right: Illustration of this effect in two dimensions. The central
panel shows the source plane with the caustic curves, the right one the image
plane with the corresponding critical curves. These curves correspond to the
red dots in the left panel. Whenever the source crosses one of the curves, two
merging images (dis)appear. At this moment, the images are magnified extremely.
}
\label{fig:lenseq}
\end{figure}

\begin{figure}[htb]
\centering
\raisebox{2.3ex}{\includegraphics[width=0.3\textwidth]{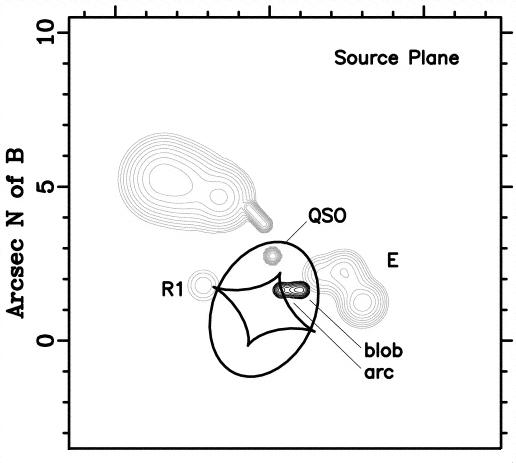}}%
\hspace{0.1\textwidth}
\includegraphics[width=0.3\textwidth]{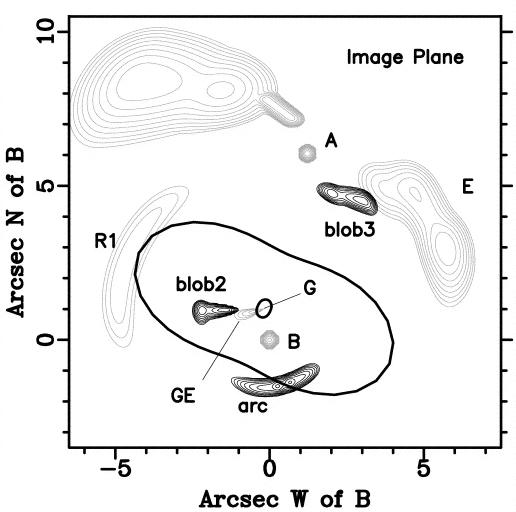}
\caption{Schematic source plane (left) and image plane (right) of
  0957+561. Shown are caustics/critical curves, radio contours in grey and
  optical contours in black. We see two images of the core but only one image
  of the extended radio jet and lobes.
\citep[from][]{avruch97}}
\label{fig:0957 max}
\end{figure}

Doubts were expressed by \citet{roberts79} when first interferometric
observations showed the complicated structure of the double radio source (see
Fig.~\ref{fig:0957 interf}). The \emph{A} image seems to show a jet and radio
lobes and one could naively expect that the same should be true for \emph{B}
if both are really lensed images of one source. In the maps, on the other hand,
image \emph{B} looks perfectly point-like.
However, Fig.~\ref{fig:lenseq} illustrates how different parts of the source
may well be lensed with different multiplicity. The specific situation for
0957+561 is shown in Fig.~\ref{fig:0957 max}.

\section{First VLBI observations of lenses}

\begin{figure}[htb]
\includegraphics[width=0.5\textwidth]{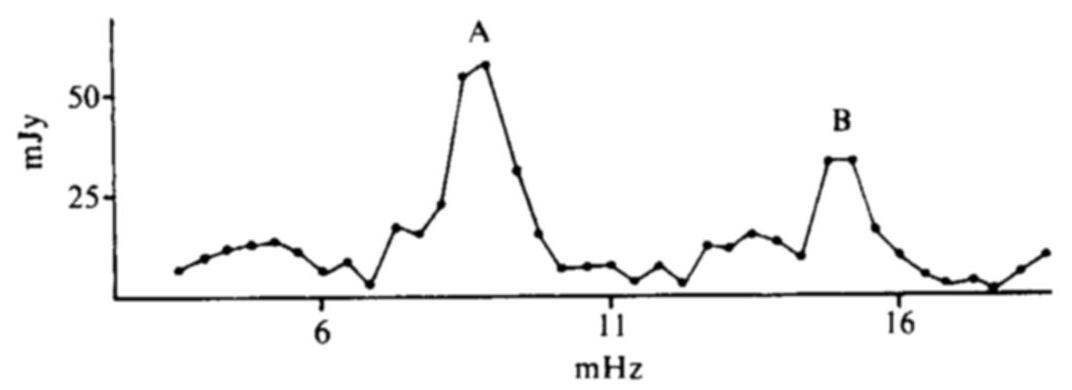}\hfill
\raisebox{2ex}{\includegraphics[width=0.25\textwidth]{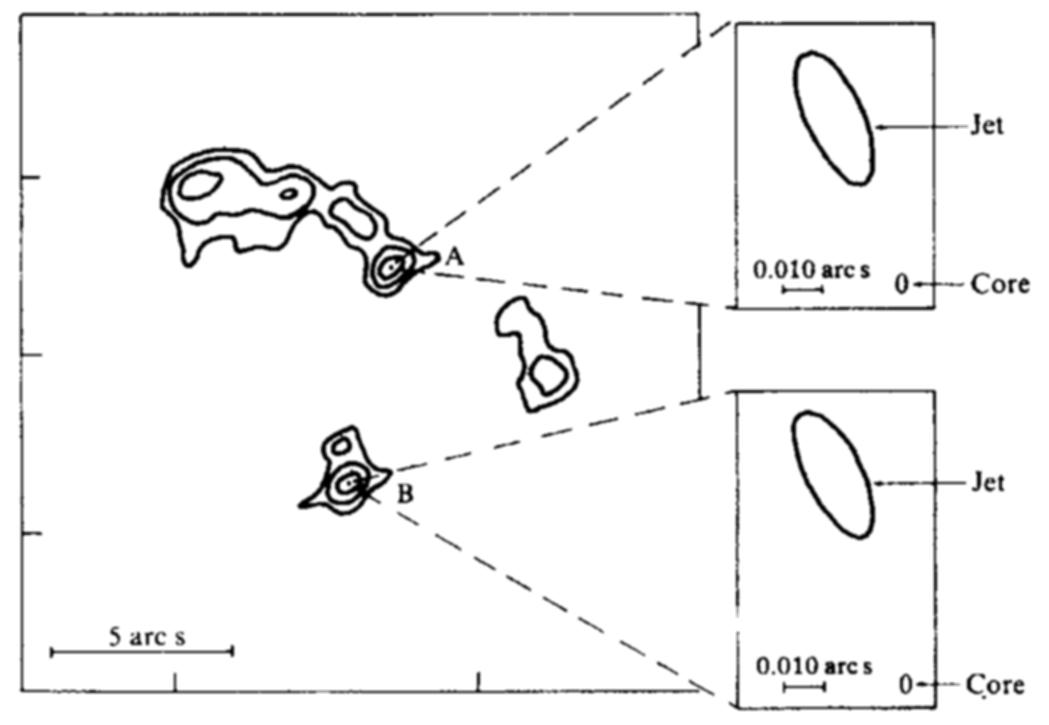}}\hfill
\raisebox{0ex}{\includegraphics[width=0.19\textwidth]{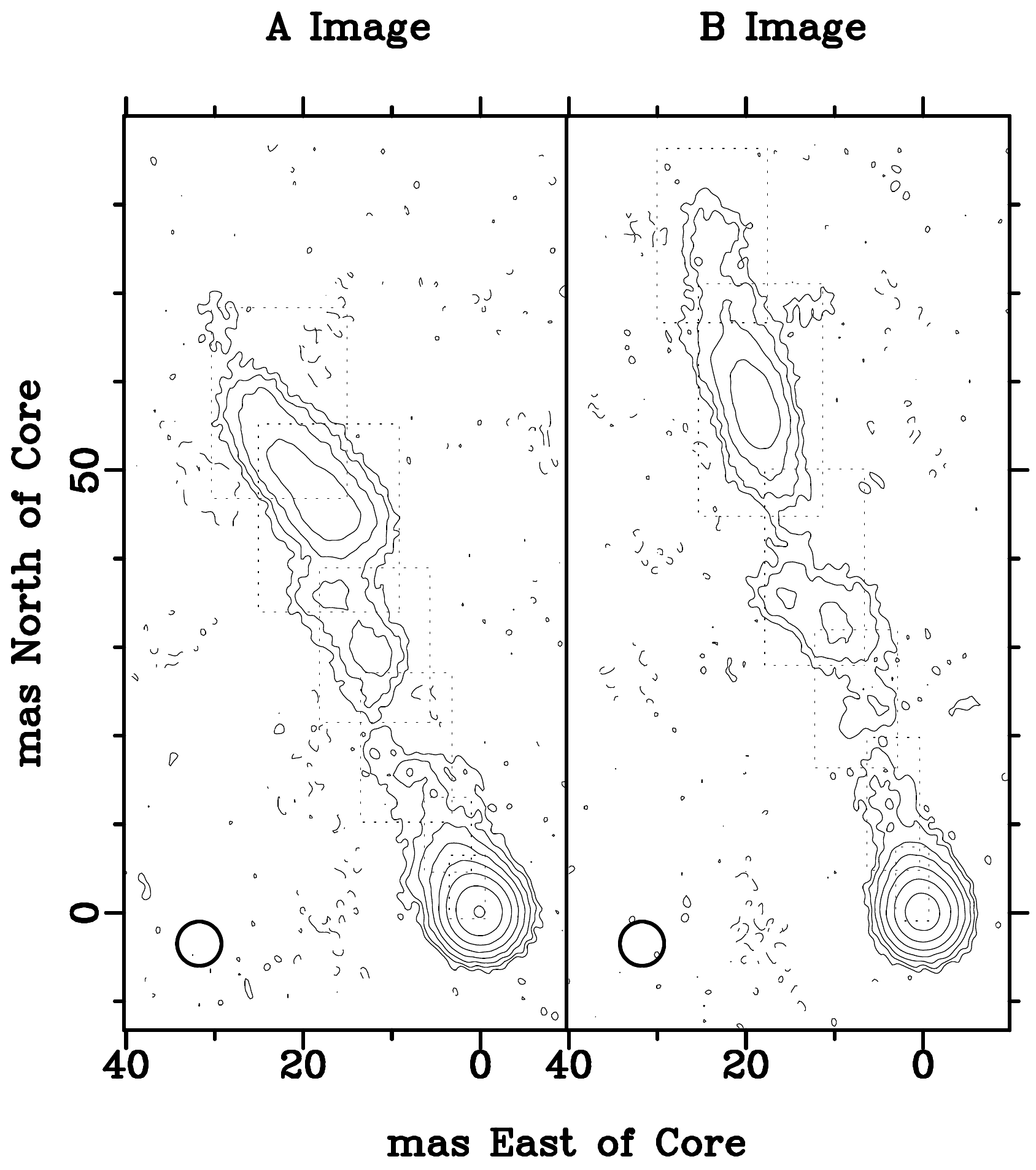}}
\caption{Left: Residual fringe rate spectrum of 0957+561 from the first VLBI
  observations; 20-min integration, Effelsberg-Jodrell baseline
  \citep[from][]{porcas79}. Centre: First Gaussian model fit results
  \citep[from][]{porcas81}. Right: Later detailed maps for comparison
  \citep[from][]{campbell95}.
}
\label{fig:0957vlbi}
\end{figure}

Naturally, the first discovered lens was also the first target for VLBI
observations (Fig.~\ref{fig:0957vlbi}). In the fringe-rate spectrum, we see
two peaks corresponding to the two images \emph{A} and \emph{B}, which thus
both must have compact structure on scales below 20\,mas. The first
multi-component model fits show that both images have a very similar
structure, consisting of a compact core component and an elongated feature
that probably corresponds to a jet. These jets show many details in VLBI maps
produced later.

It is impressive to see how far VLBI has
evolved from simple fringe-rate spectrum plots to detailed maps of
gravitational lenses. The first observations also show that real science can
be done with only one VLBI baseline.

\section{Fields of study}

\begin{figure}[htb]
\centering
\ifpdf%
\includegraphics[angle=-90,width=0.8\textwidth]{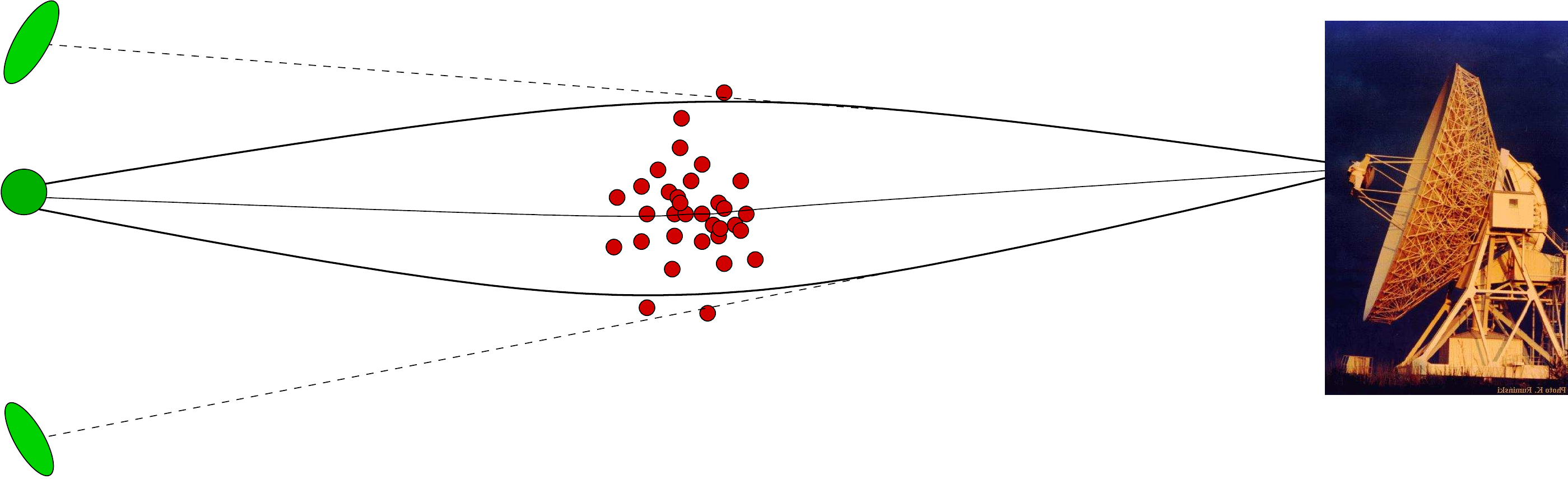}
\else%
\includegraphics[width=0.8\textwidth]{diag2}
\fi
\caption{Lensing can be used to study the lensed sources (left), the lenses
  themselves (centre), propagation effects and extinction (anywhere between
  source and observer), and the properties of global spacetime (including
  cosmology and tests of relativity.)}
\label{fig:fields}
\end{figure}

After this introduction, we come to a more systematic discussion of the several
topics that can be studied with gravitational lensing. Fig.~\ref{fig:fields}
schematically shows a typical lensing scenario with multiple images of one source.

\subsection{Lenses as natural telescopes}

\begin{figure}[htb]
\includegraphics[height=0.35\textwidth]{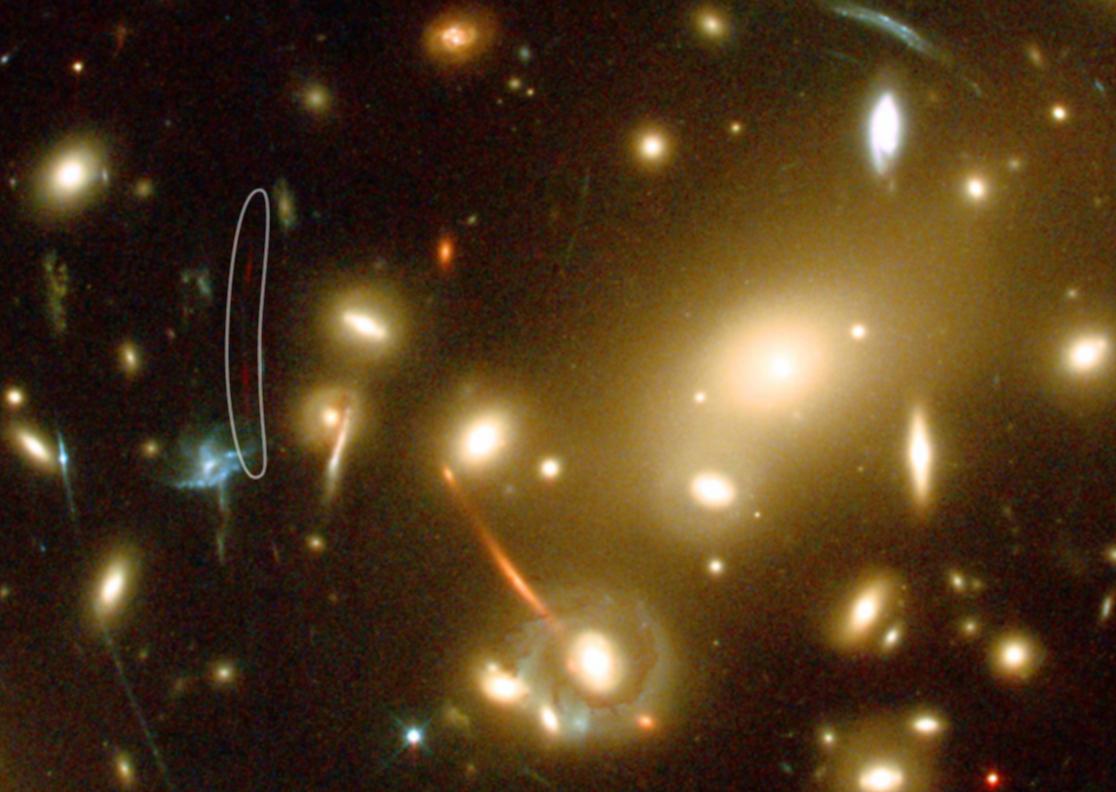}\hfill%
\includegraphics[height=0.35\textwidth]{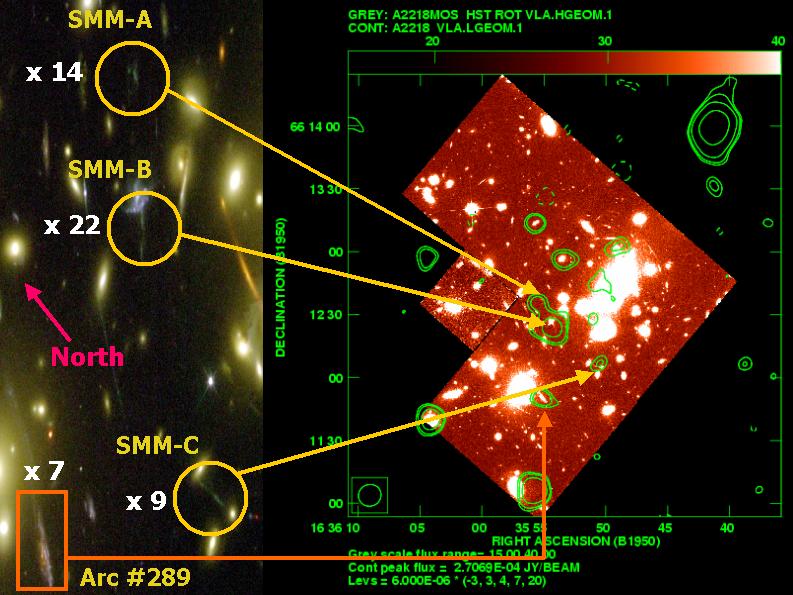}
\caption{The lensing cluster of galaxies A2218. Left: A $z=7$ background
  galaxy amplified by a factor of $\mu\approx 25$ \citep[from][]{kneib04}.
Right: The first radio detection of a lensed star-forming galaxy \citep[from][]{garrett05}.}
\label{fig:a2218}
\end{figure}

\begin{figure}[htb]
\centering
\includegraphics[width=0.6\textwidth]{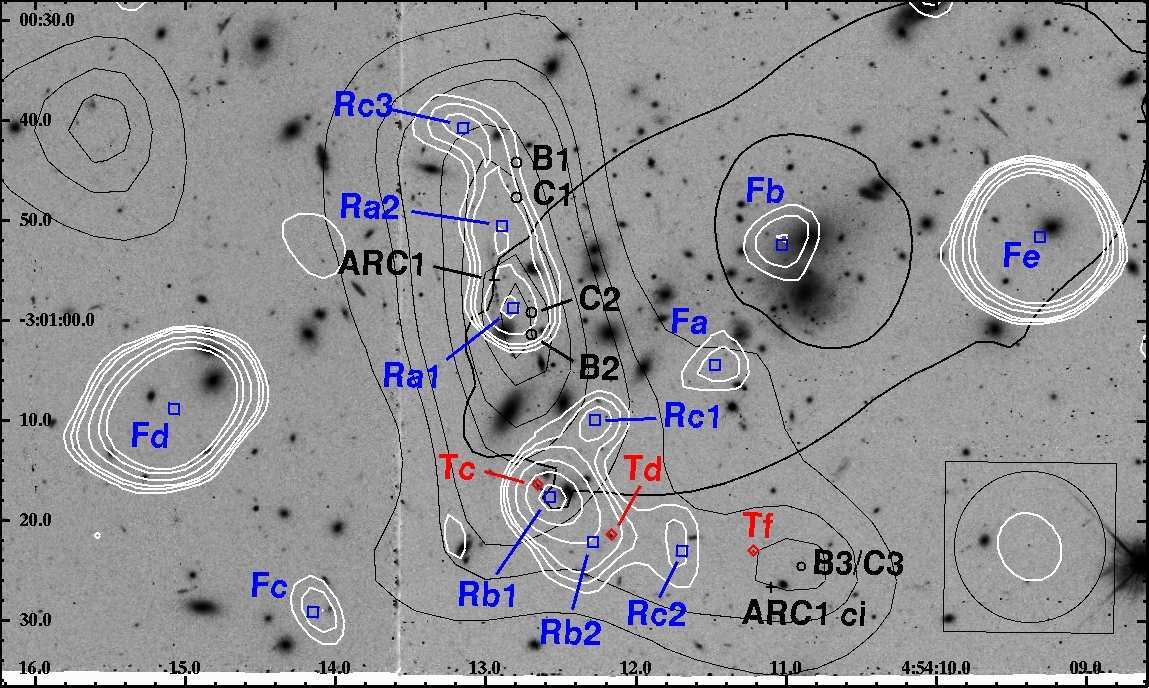}
\caption{The cluster MS0451.6--0305 shows a number of lensed background source
  components in the radio (white contours) and sub-mm (black contours) domains. Lens
  models predict that they all form a compact ensemble of forming
  or merging galaxies at high redshift. \citep[from][]{berciano06}}
\label{fig:alicia}
\end{figure}

The lens effect can be utilised to extend the capabilities of our instruments
because it provides additional magnification and corresponding flux
amplification, and in this way boosts the effective resolution and sensitivity
of observations.
Clusters of galaxies provide the largest lensing cross-sections and are thus
the primary of these \emph{natural telescopes}. Two examples are shown in
Figs.~\ref{fig:a2218} and \ref{fig:alicia}. In one case, the amplification made
it possible to detect a distant galaxy at $z=7$ and observe its optical
spectrum. In the others, the lensing amplification enhances weak, background
star-forming galaxies that have first been found at sub-mm wavelength
above the radio detection limit. Without the lenses, current radio telescopes
would not be sufficiently sensitive to study these objects at all.

In galaxy-scale lenses, the regions of high amplifications are very small, but
in rare occasions it happens that a radio source is located so close to a
caustic that it is amplified by factors of a few hundred. B2016+112 is one of
those examples. In another talk at this meeting, \cite{more06} present new HSA
observations of 
this system, showing source components almost merging at the critical curve
with extreme magnifications. Back-projecting these components into the source
plane will in the future provide $\mu$arcsec-resolution maps of this source.

\subsection{Mass distributions of lenses}

Studying the mass distributions of lenses is one of the most important fields
in lensing research for several reasons. On one hand, the mass distribution
governs the light deflection and image configuration, including
magnifications. Without a good knowledge of these properties, applications of
lenses as natural telescopes (or for other purposes) would become inaccurate
and speculative.

However, this argument can be turned around and we can use the image
configuration to constrain the mass distribution of the lens.
This method provides the most accurate information about the mass distributions
of very distant lenses, allowing the study of structure and evolution of
galaxies in a systematic way.
The standard approach to do this is to assume a simple parametrised
mass distribution for the lens and make simple assumptions for the source
structure, e.g.\ assume that it consists of a small number of compact
components.
The model parameters (of the lens and source) are then varied in a way to fit
the predicted image configuration to the observed one.
This can include image positions as well as shapes of compact but slightly
extended components.
Generally, it is advantageous to have many lensed source components, but they
have to be of a sufficiently simple structure to make this standard approach
viable.

\begin{figure}[htb]
\centering
\includegraphics[height=0.35\textwidth]{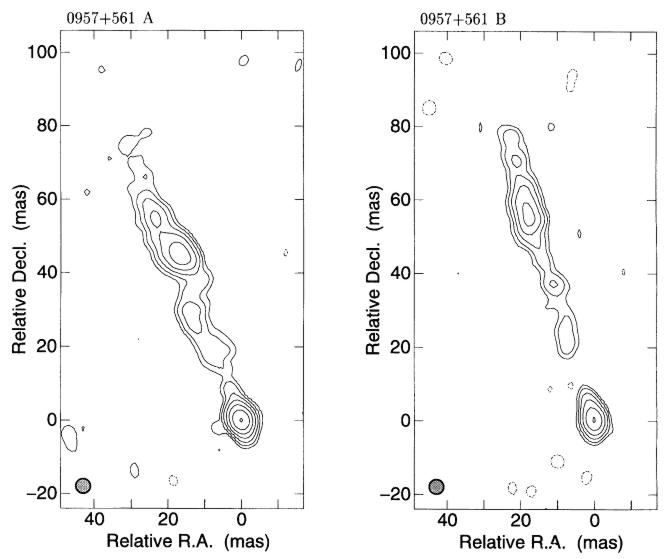}\hspace{0.07\textwidth}
\includegraphics[height=0.35\textwidth]{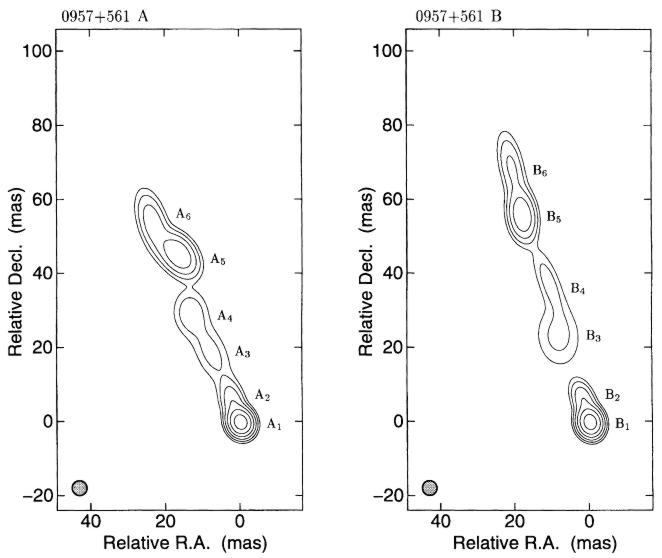}
\caption{Constraints from the substructure of both jet images in 0957+561. The
  observed structure (left panels) of the jet images is parametrised as a
  collection of Gaussian model components (right panels). The parameters of
  these components are then used to determine a relative magnification
  matrix that linearly maps the central region of image \emph{A} to \emph{B}
  or vice versa. The mass-model parameters are then varied to fit the
  predicted matrix to the one determined from the observations. \citep[from][]{garrett94}}
\label{fig:model 0957}
\end{figure}

Fig.~\ref{fig:model 0957} illustrates this for the case of 0957+561,
where the source is complicated enough to make this procedure a non-trivial
task. The constraints are quite good, but unfortunately the structure of the
lens (which consists of a massive cD galaxy together with its surrounding
cluster) is so complicated that even these constraints are not sufficient to
provide a good accuracy for the lens mass distribution. One
disadvantage is that all the jet components are located very close to each
other so that they sample the light deflection only in small regions,
basically providing the deflection angle and its derivatives at only two
points.

\begin{figure}[htb]
\centering
\includegraphics[width=0.75\textwidth]{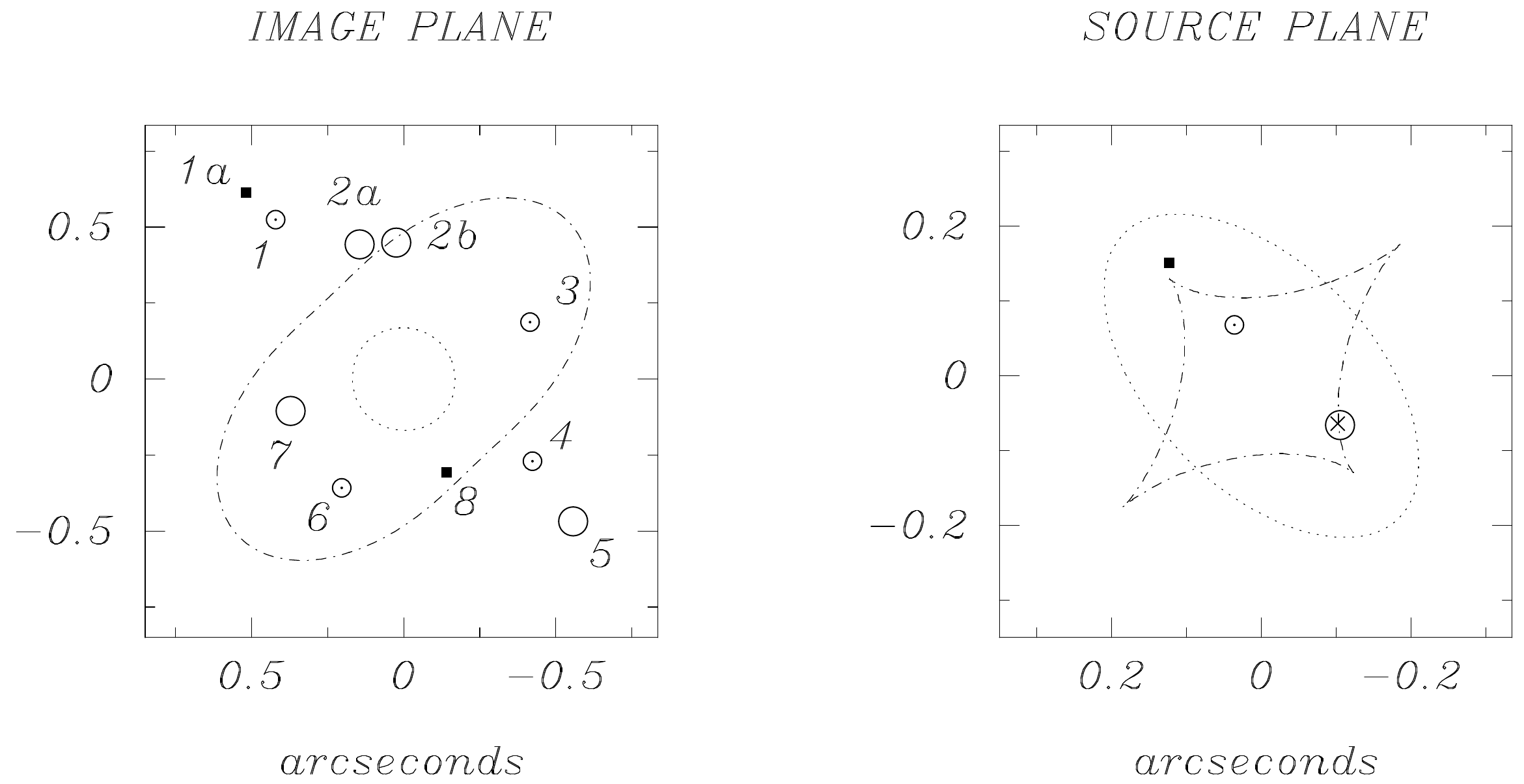}
\caption{The lens B1933+503. Left: Image configuration with critical curves. Right: Source
  component configuration with caustics. One component (black dot) is
  doubly imaged, the other two (circles) are both quadruply imaged,
  resulting in ten images of these three components. The images probe the
  lensing potential at a wide range of radii. \citep[from][]{nair98}}
\label{fig:1933}
\end{figure}

In order to obtain more information about the global properties of the mass
distribution, it is necessary to have images at widely separated positions,
especially at different distances from the lens centre. B1933+503 is a unique
lens system consisting of \emph{ten} images of several source
components (see Fig.~\ref{fig:1933}). Parametrised model fits  \citep{cohn01} for this system
show that the mass distribution must be very close to isothermal (projected
surface mass density $\sigma\propto 1/r$), a property found in most accurately
modelled lenses.
Later on, we will discuss how to use lensed general \emph{extended} sources to
constrain lens models.

\begin{figure}[hbt]
\centering
\begin{minipage}{0.23\textwidth}
\includegraphics[width=\textwidth]{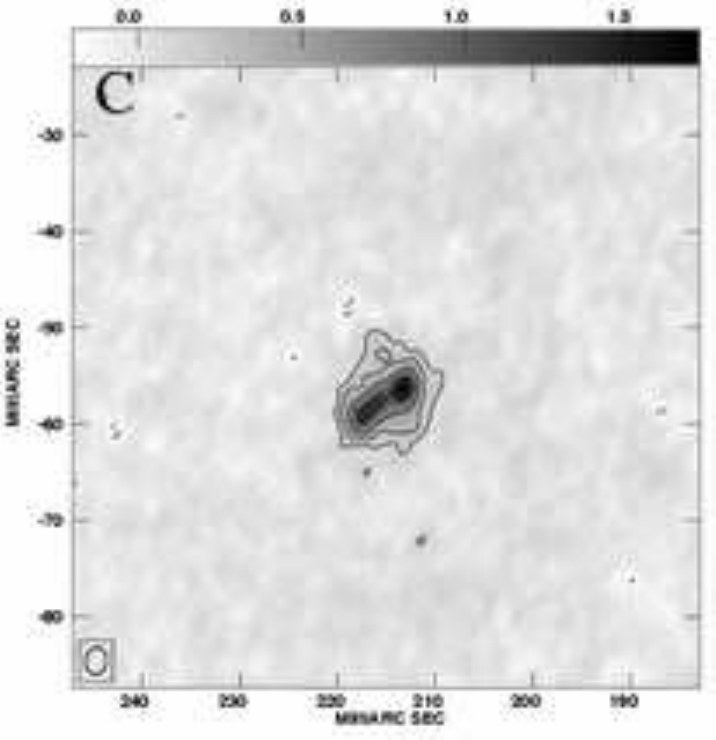}\\
\includegraphics[width=\textwidth]{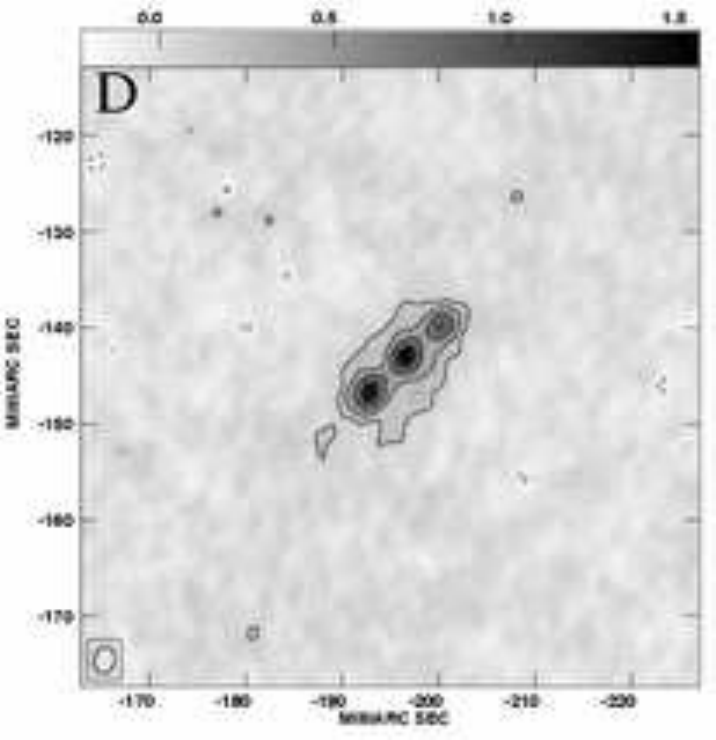}
\end{minipage}%
\raisebox{-1ex}{\begin{minipage}{0.5\textwidth}
\includegraphics[angle=-90,width=\textwidth]{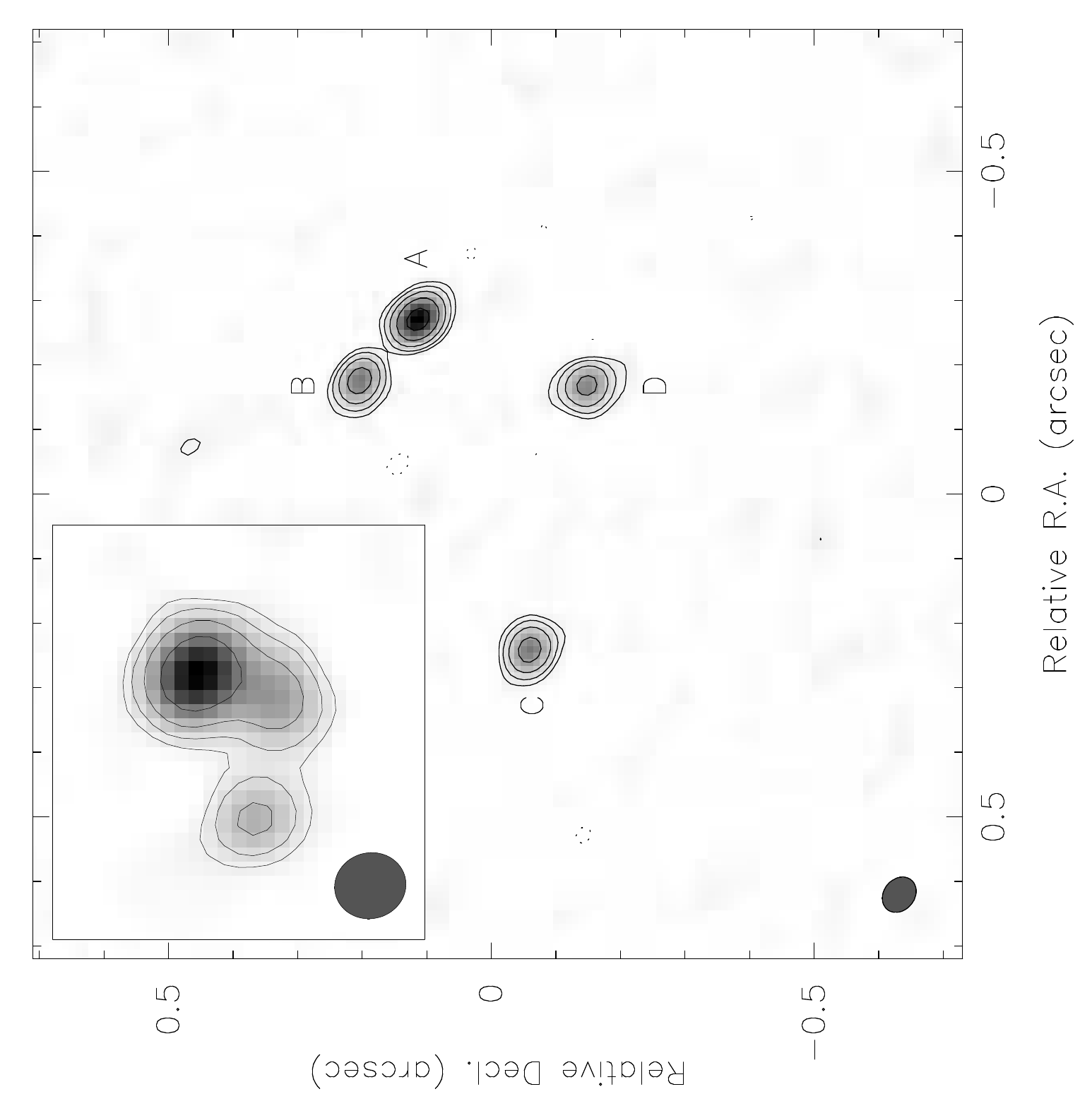}
\end{minipage}}%
\begin{minipage}{0.23\textwidth}
\includegraphics[width=\textwidth]{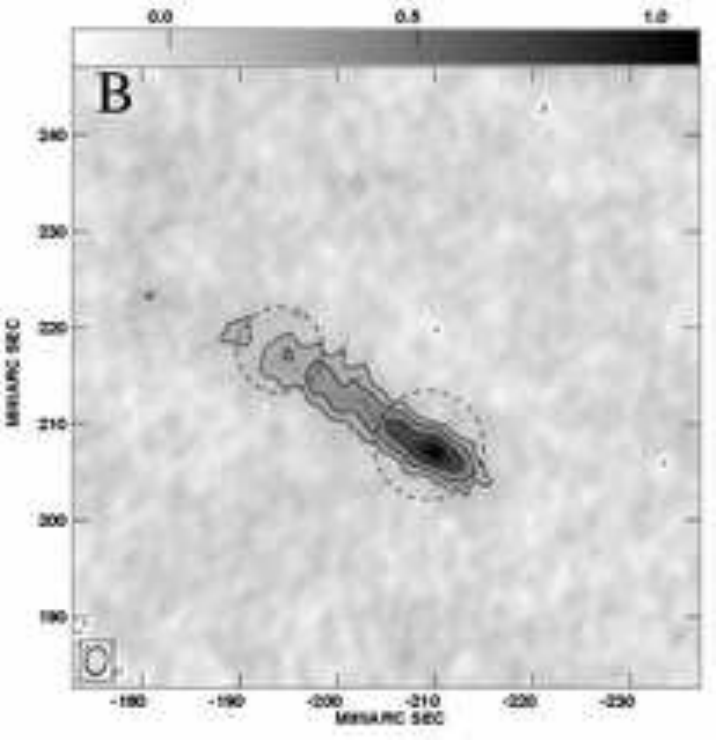}\\
\includegraphics[width=\textwidth]{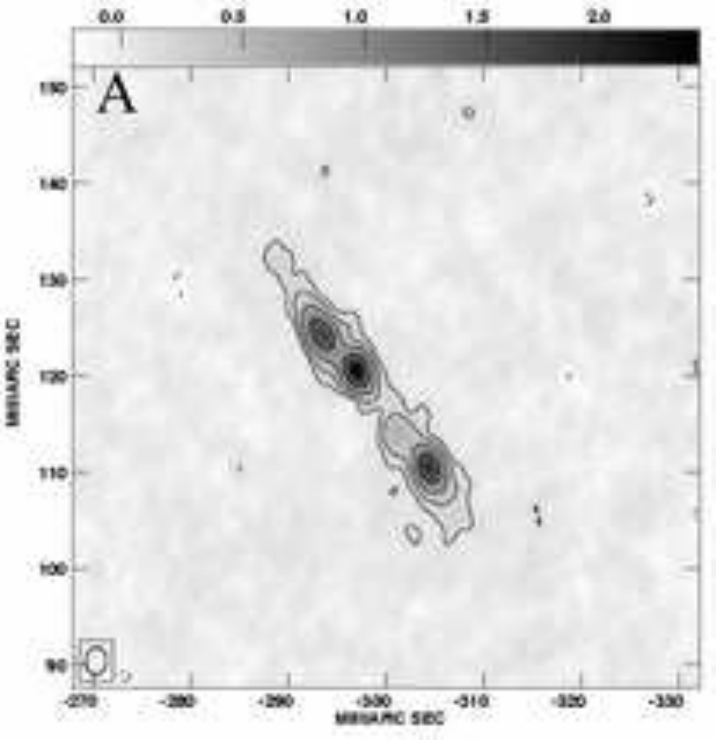}
\end{minipage}
\caption{The quadruple lens system B0128+437. Centre: MERLIN map of the
  complete system \citep[from][]{phillips00}. Outer fields: High-resolution
  VLBI maps of the four images \citep[from][]{biggs04}.}
\label{fig:0128}
\end{figure}

The lensed image configuration not only provides information about the global
mass distribution but is also sensitive for small-scale deviations as
predicted by CDM structure-formation scenarios. Fig.~\ref{fig:0128} shows an
example where each of the four images consists of three subcomponents. The
general four-image geometry can be fitted well with smooth mass distributions,
but these models cannot explain the relative positions of all subcomponents in
the images. An easy explanation would be the presence of a small-mass clump
close to one of the images that distorts the geometry very locally without
affecting the global configuration.  There are other cases where it is not the
image configuration but the flux density ratios that are in conflict with
simple models. This is not surprising since small-scale mass substructure
influences derivatives of deflection (determining the amplifications) much
stronger than the deflections (and image positions) themselves.

\begin{figure}[htb]
\includegraphics[width=0.6\textwidth]{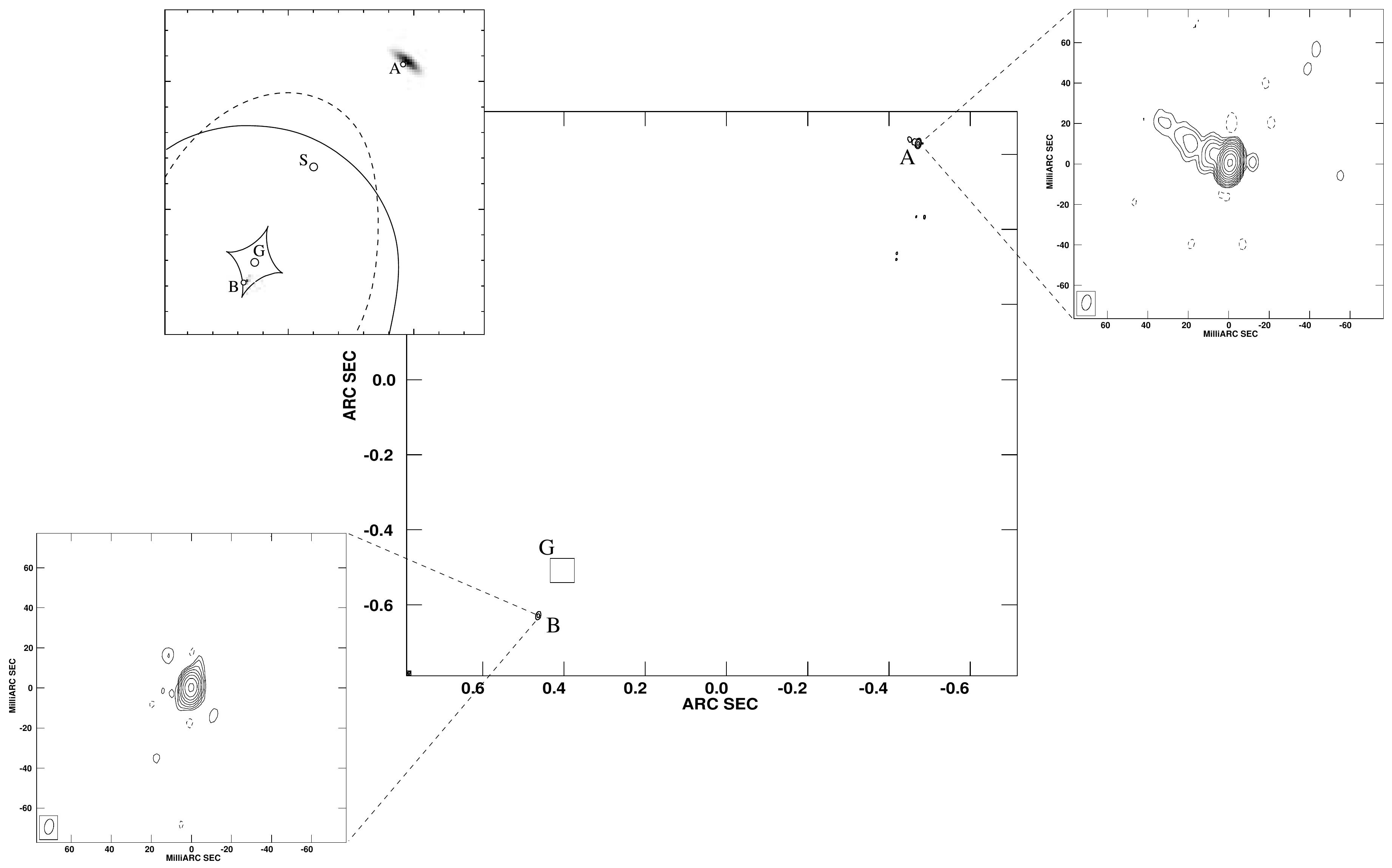}\hspace{0.035\textwidth}%
\raisebox{1ex}{\includegraphics[width=0.35\textwidth]{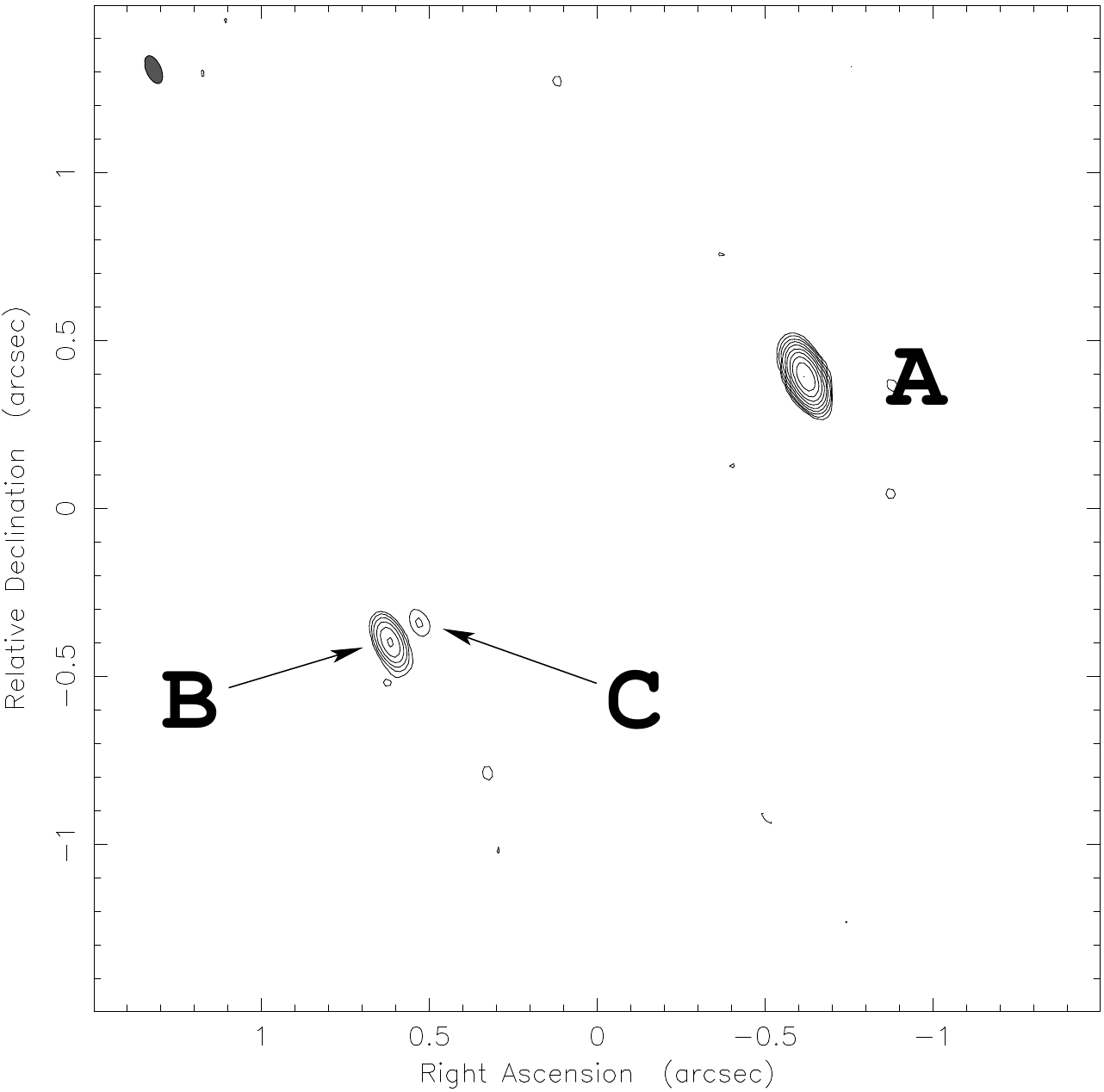}}
\caption{Two cases of studies concerning possible central images.
Left: B1030+074, where even with HSA observations no central image could be
found \citep[from][]{zhang06}. Right: J1632--0033, the currently most (and
maybe only) convincing case with a weak central image \citep[from][]{winn03}.}
\label{fig:central}
\end{figure}

Another aspect of the small-scale mass distribution that can be studied with
lensing is the mass profile in the very centres of lensing galaxies.  For a
smooth mass profile with a non-diverging deflection angle, one would always
expect an odd number of images, one of which would be located close to the
lens centre. However, if the central mass concentration becomes very steep or
even singular, this central image will be highly de-magnified or even
completely suppressed.  Currently, only one such central image is believed to
be detected (see Fig.~\ref{fig:central}, together with a case with no central
image). Instead, there is evidence that isothermality of the mass profile
typically extends very close to the lens galaxy centres, which would suppress
the central images. There is no good explanation for this fact in standard
structure-formation theories.

\subsection{Propagation effects: scattering, absorption etc.}

The study of propagation effects like scatter broadening, free-free absorption
or extinction and reddening in the optical domain is usually plagued by the
lack of knowledge of the \emph{intrinsic} structure and spectral energy
distribution of a distant source, with which the \emph{observed} properties
could be compared.  Gravitational lensing kindly produces two or more copies
of the same source component that are identical in their spectra and have
only simple (and easy to model) differences in their total flux density and
source structure. By comparing properties of these images, one can directly
infer the \emph{differential} propagation effects like differential reddening
or differential scatter broadening, without assumptions about the intrinsic
properties.  This will be shown for one example later.

\subsection{Measurement of distances and cosmology}

Long before the discovery of the first lens, it was argued by 
\citet{refsdal64b,refsdal66} that the lens effect can be used to test
cosmological theories and to measure distances and thus the Hubble
constant in particular.
The principle behind this idea is very simple. In a typical lens
system, we can measure the angular separations between the observed images and
the lens. Modelling the mass distribution then gives us all the other angles
defining the geometry (like the true source position). Ratios of distances
between observer, lens and source can be derived easily from the observed
redshifts. We conclude that in such a situation the complete lensing
geometry is known, \emph{except for the scale.} If only one length in the
system can be measured, all other lengths, including the distances to the lens
and source, can immediately be calculated.
It was the idea of \citet{refsdal64b} that the light travel-times will differ
from image to image (because of different geometrical paths and different
Shapiro delays) and that this light travel-time difference can be used as the
defining length to scale the whole geometry. If the background source is
variable, these variations will be seen in all the images, but shifted
relative to each other by the corresponding time delays.
Once the distances are determined, they can be used together with the
redshifts to estimate the Hubble constant. Distances for given redshifts are
inversely proportional to the Hubble constant, and the time delay is
proportional to the distances, so that the product $H_0\Delta t$ is a constant
that can be derived from the lens mass model, which, in turn, is constrained by
the image geometry.
This method has been applied to a number of lens systems, leading to more or
less consistent results. We will discuss only one example further below.

\subsection{Tests of relativity}

The most fundamental test of relativity performed with the lens effect
consists of measuring the deflection angle in our solar system and comparing
it with the theoretical expectations.
For this purpose the deflection can be parametrised as
\[ \alpha = 2(1+\gamma)\frac{GM}{c^2 r} \]
and the resulting $\gamma$ can be measured. Two special values would be
$\gamma=0$ as expected from Newtonian theory and $\gamma=1$ as expected from
general relativity. The work of \citet{eddington19} was only a first example,
and the accuracy has improved by several orders of magnitude by using VLBI
techniques.
\citet{shapiro04} found $\gamma=0.9998\pm0.0004$ from a combination of very
many geodetic VLBI data sets, which confirms general relativity with high
accuracy.

It has also been claimed that the deflection of light by a moving object
(Jupiter in this case) can be used to measure the propagation speed of gravity
\citep{kopeikin01}. Such an experiment has actually been carried out
\citep{fomalont03}, and the outcome is consistent with relativity. However,
there are serious arguments against this interpretation of the experiment
\citep[see e.g.][]{will03,carlip04} and the debate has not led to an agreement
between the opponents yet.

\section{The lens B0218+357}

This lens system is one of the best studied
cases and is a good example for many of the applications of the lens effect
discussed above. It consists of two bright images and an additional Einstein
ring with the same diameter as the image separation (Fig.~\ref{fig:0218
  colour}). The two images are resolved by VLBI and show the core and inner
jet of the lensed background source \citep[e.g.][]{biggs03}.

\begin{figure}[htb]
\includegraphics[width=0.49\textwidth]{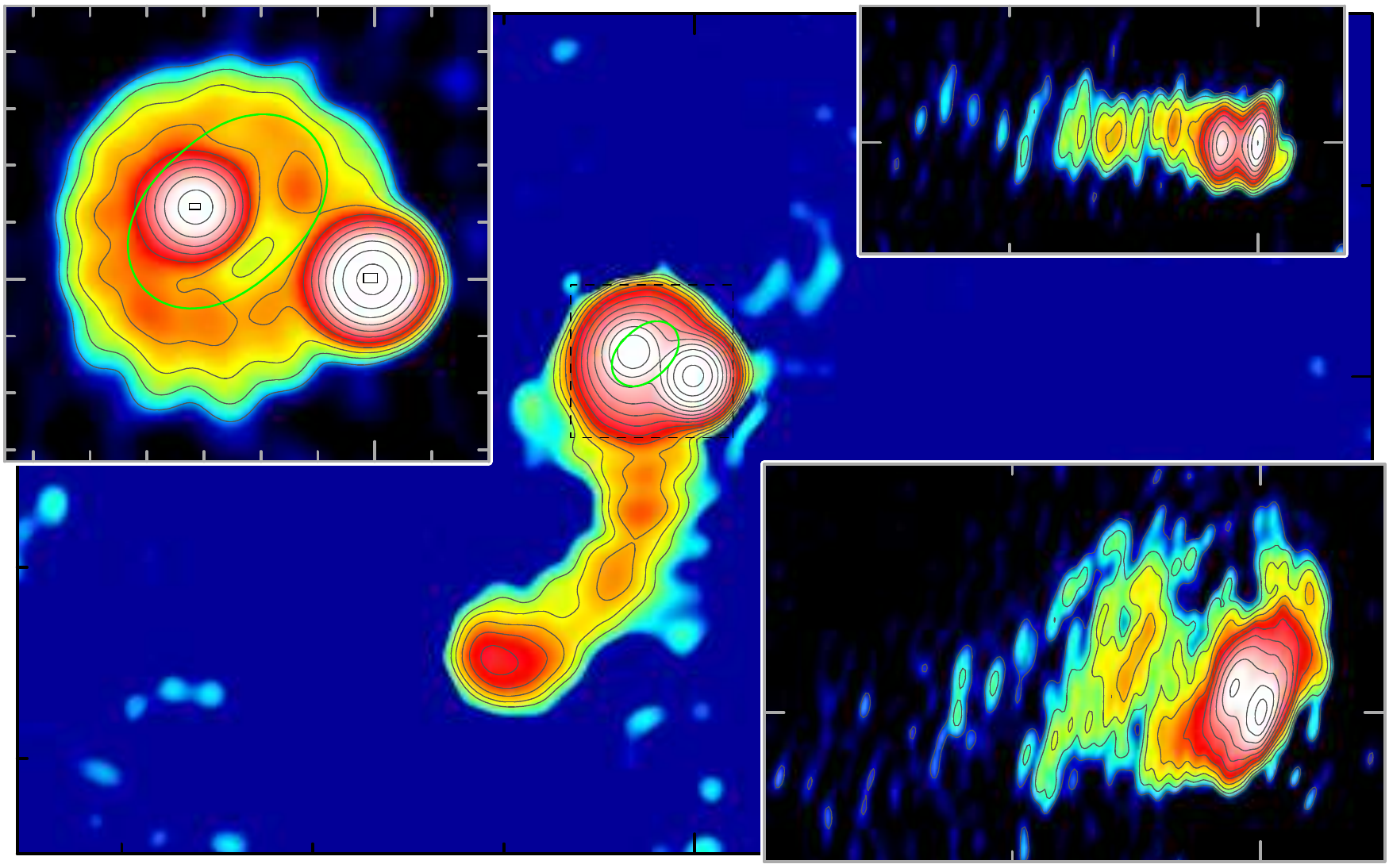}\hfill%
\includegraphics[width=0.49\textwidth]{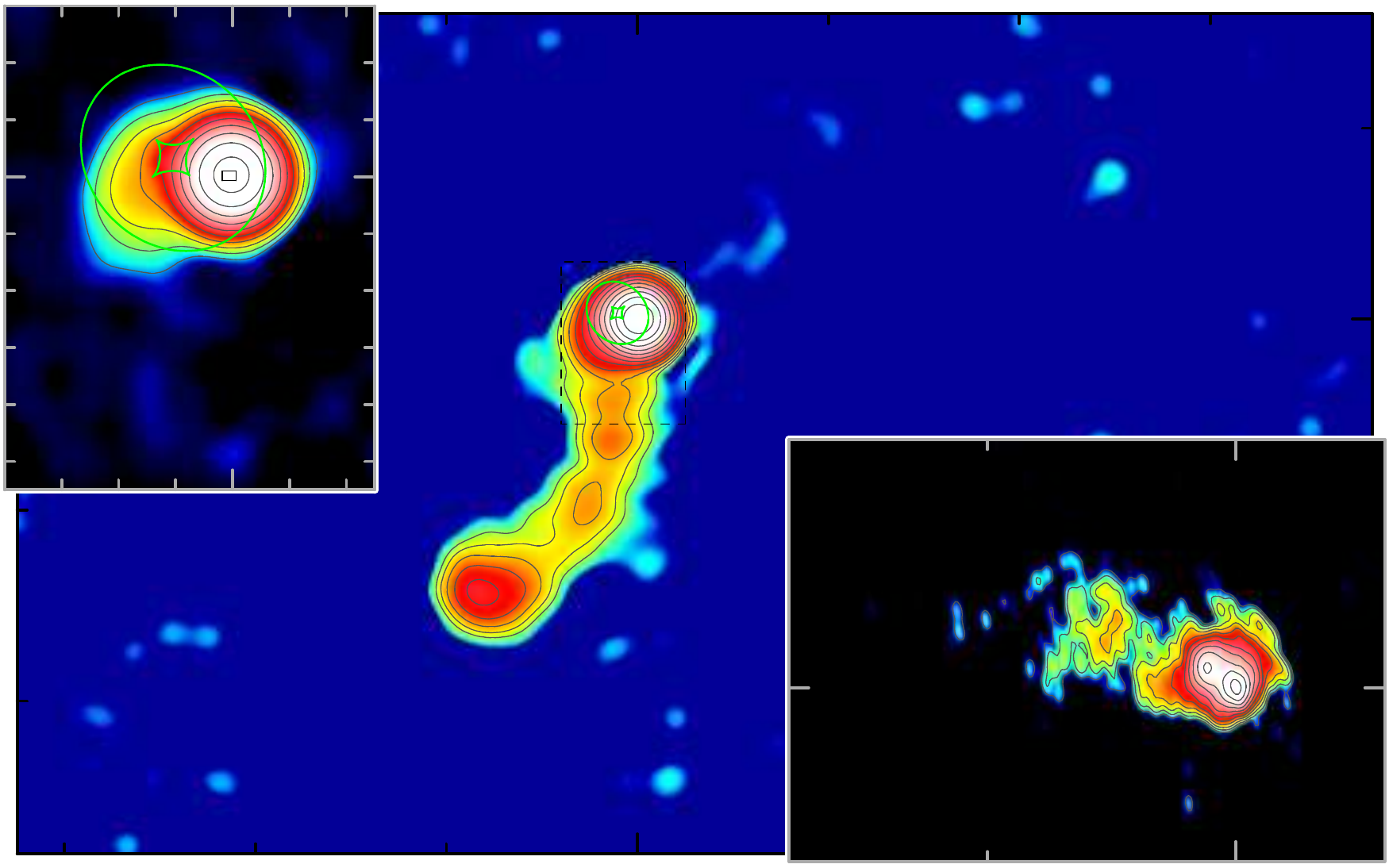}
\caption{The lens B0218+357. Left: The lens plane as we see it. The central
  map provides an overview. The Einstein ring and the two images are shown
  in the upper left; the insets at the right show the magnified VLBI
  substructure of the two images. The magnified regions are marked by black
  rectangles. Right: The reconstructed (unlensed) source
  plane. \citep[from][]{lc2}
}
\label{fig:0218 colour}
\end{figure}

\subsection{Lens models and Hubble constant}

B0218+357 has some unique properties that make it particularly well suited to
determine the Hubble constant using Refsdal's method. Firstly, the time delay
is known with sufficient accuracy. \citet{biggs99} find $\Delta
T=(10.5\pm0.4)\,\rm d$, a result that is consistent with that of
\citet{cohen00}.  Secondly, the lens is an isolated spiral galaxy without
close neighbours or clusters nearby, which allows the use of simple models
with a small number of parameters as a realistic description. Finally, the
wealth of structure in the Einstein ring and the VLBI maps of the lensed jets
provides a good number of valuable constraints for the modelling. The VLBI
substructure is especially sensitive to the radial mass profile, which seems
to be very close to isothermal but slightly shallower \citep{biggs03}.
Unfortunately, the accurate lens position (relative to the images) was not
known until recently. The image separation is only $330\,\rm mas$, the
smallest of all galaxy lenses, which makes direct optical measurements very
difficult.  However, the structure of the Einstein ring can be used to
determine the lens position indirectly. Contrary to the lens modelling methods
described before, the source cannot be described as a small number of
point-like sources anymore. Instead, more sophisticated methods must be used,
in which the true source structure is also fitted for in a non-parametric way.

We used our own improved version of the LensClean algorithm to accomplish this
task \citep{lc1}.  LensClean iteratively builds a source model similarly to
the normal Clean algorithm, but takes into account the effect of the (for the
moment fixed) lens model. `Clean components' are allowed only in combinations
that are consistent with the effect of the lens. If, for example, a component
is to be included at a position that is quadruply imaged, all four lensed
images have to be included in the model, scaled with their respective
amplifications.  Once converged, this inner loop of LensClean has determined a
source model that minimises the residuals given a certain lens model. An outer
loop then varies the lens model parameters, again in order to minimise the
residuals.  The final result is a simultaneous fit of lens and source model.
The main result in the case of B0218+357 is the lens position, which then
directly translates to a value for the Hubble constant \citep{lc2}:
\[
H_0 = (78\pm6) \;\rm km\,s^{-1}\,Mpc^{-1}
\]
The lens position was later confirmed by a direct optical measurement using
the HST/ACS \citep{york05}.  The result for $H_0$ is consistent with other
methods that use completely independent information. Because it is determined
by a direct one-step method, the systematic uncertainties inherent in the
complex distance-ladder methods are avoided.

\subsection{Propagation effects}

\begin{figure}[htb]
\includegraphics[height=0.28\textwidth,clip]{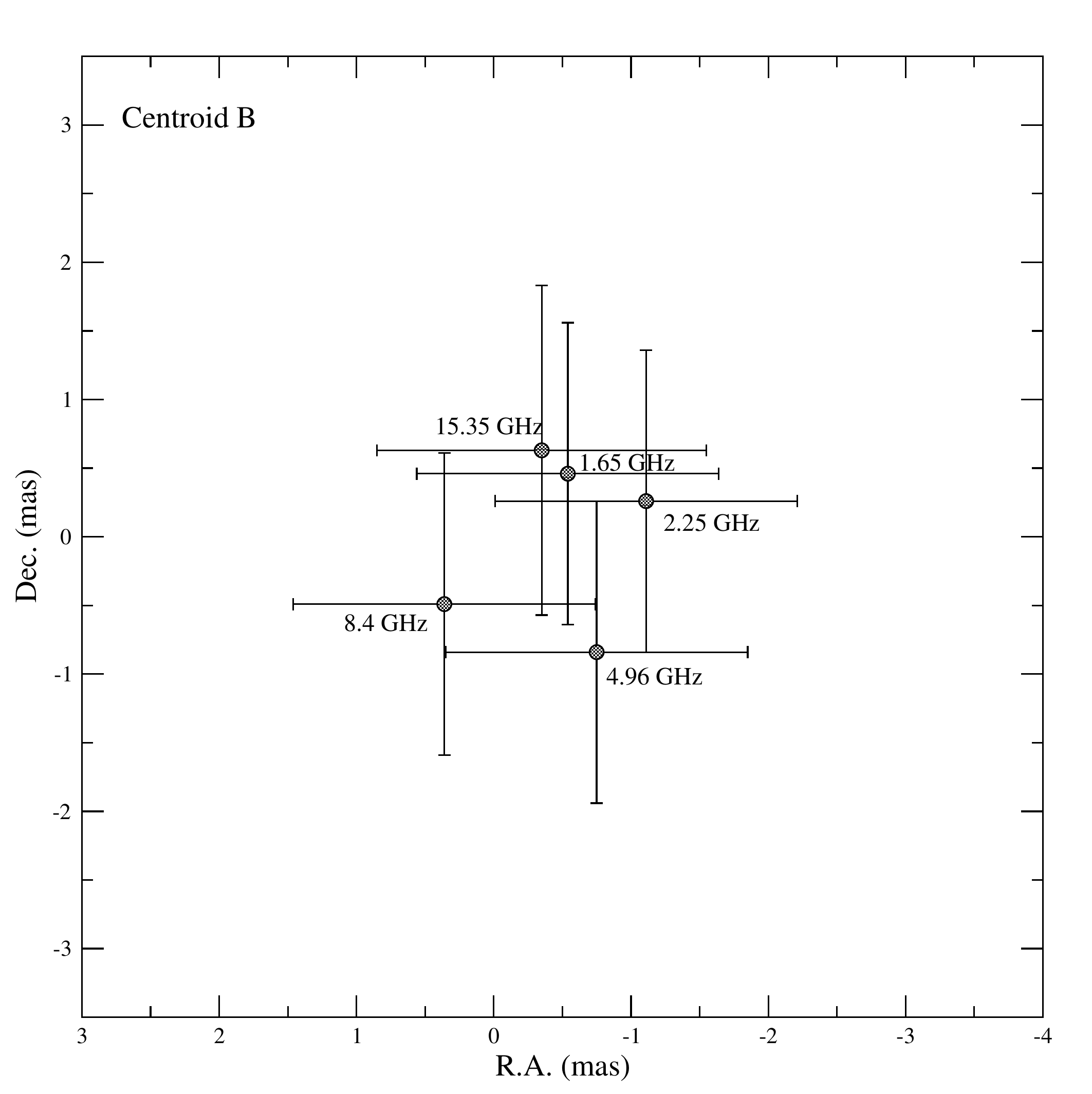}%
\includegraphics[height=0.28\textwidth,clip]{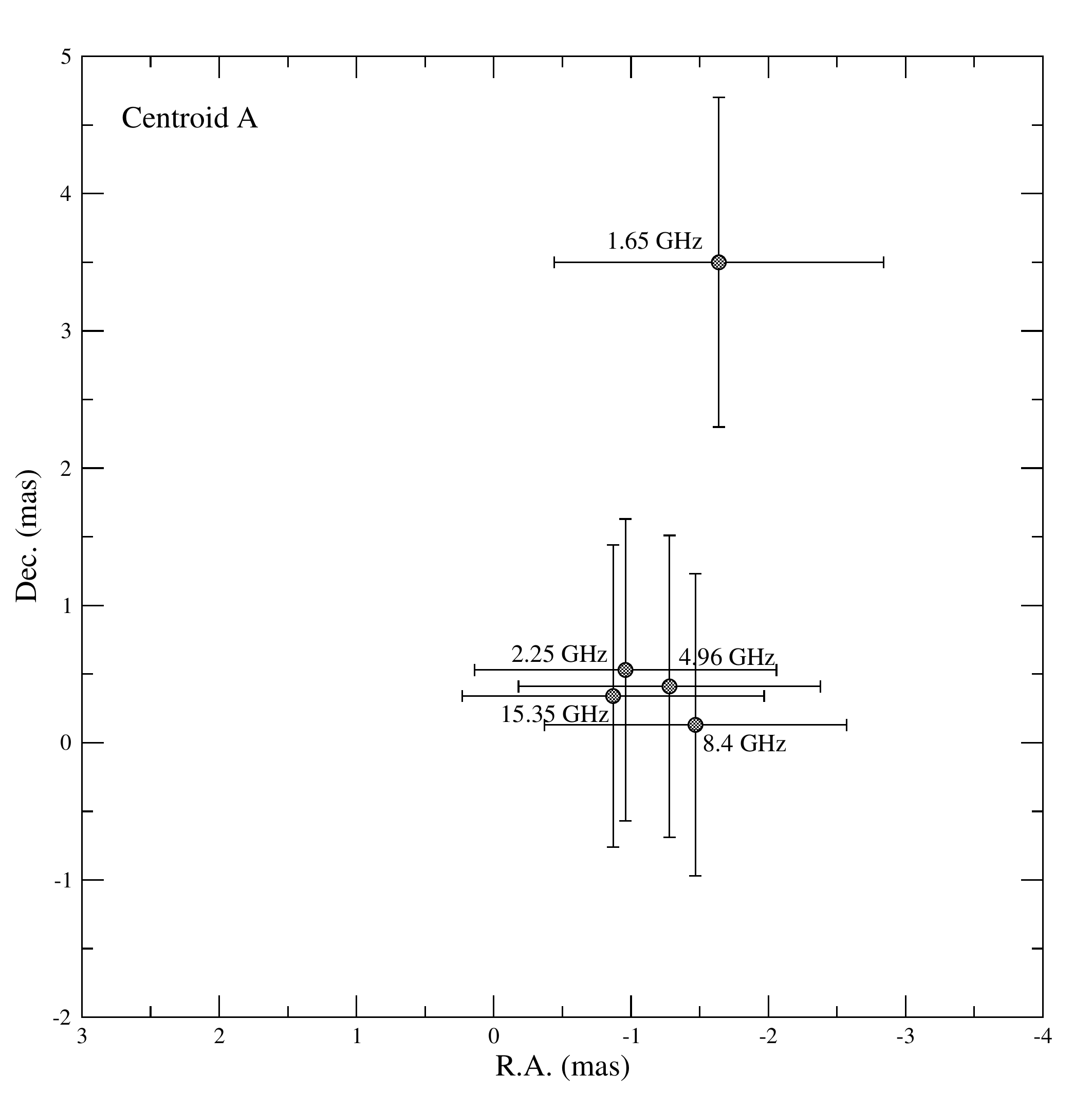}\hfill
\includegraphics[height=0.28\textwidth]{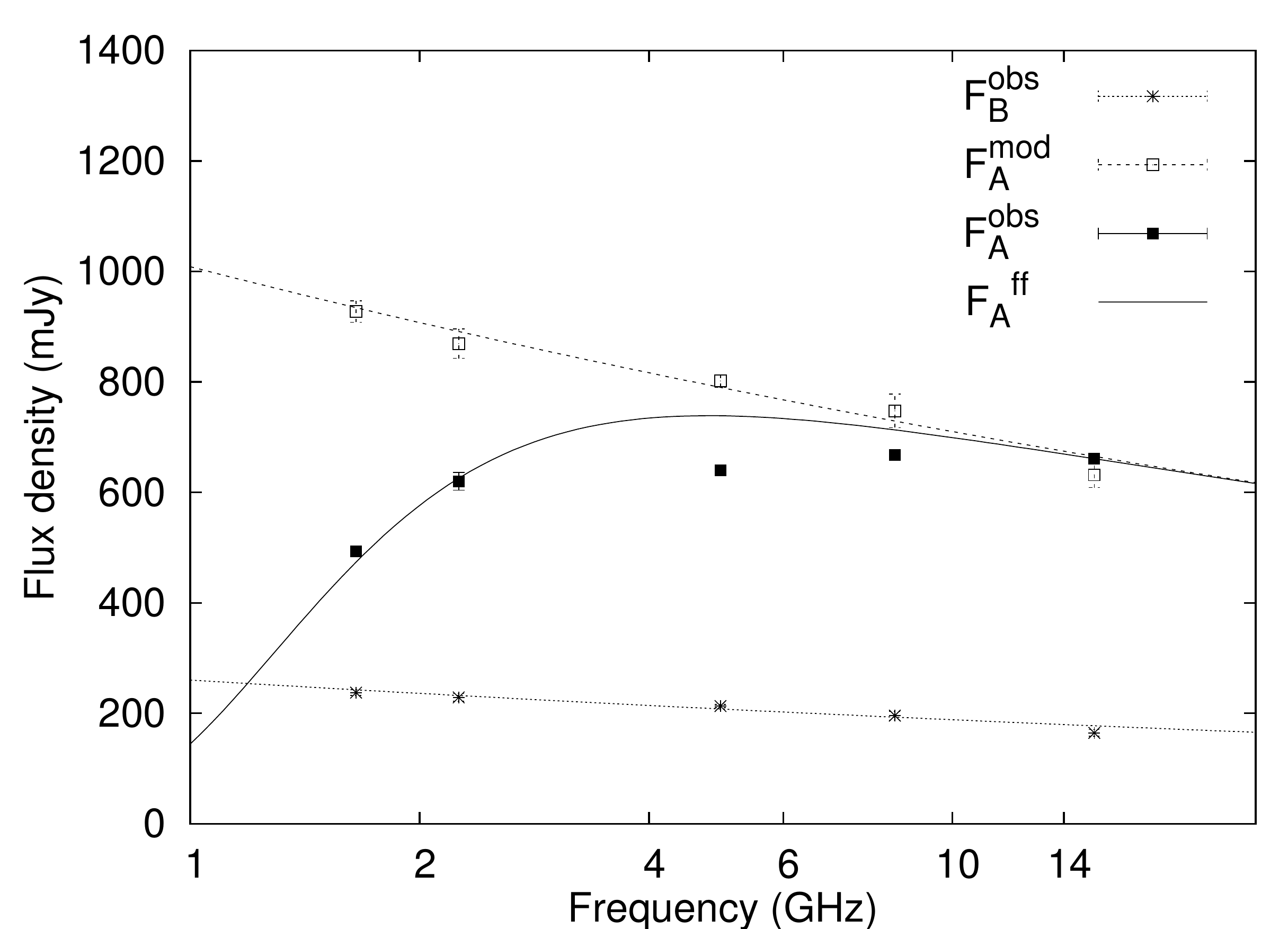}
\caption{Left: Image positions of $A$ and $B$ for different frequencies
  \citep[from][]{mittal06}. Right: free-free absorption model for the flux
  density of $A$ derived from $B$ ($F_A^{\rm ff}$) with corresponding
  measurements ($F_A^{\rm obs}$). The agreement is not perfect but reasonably
  good \citep[from][]{mittal06b}.}
\label{fig:mittal}
\end{figure}

It has been known for long that the flux density ratio of the two images $A/B$
shows a strong frequency dependence. At high frequencies, it is close to 4 and
decreases for lower frequencies.  This is surprising since lensing is
achromatic and should not change the spectra of the images. One possible
explanation would be source shifts as a function of frequency, which, together
with the strong magnification gradient, could explain the observed trend,
provided they were of the order $10\,\rm mas$ or more.  \citet{mittal06}
investigated this possibility by measuring the image positions with
multi-frequency, phase-referencing VLBI observations. They found (see
Fig.~\ref{fig:mittal}) that any image shifts are of the order of $1\,\rm mas$
or less, not enough to explain the observed effect. Similarly, changes of the
source structure could be ruled out as reason for the changing flux density
ratios.  The most plausible explanation is a free-free absorption, mainly in
the \emph{A} image.

\subsection{More VLBI aspects}

\begin{figure}[htb]
\begin{minipage}{0.38\textwidth}
\includegraphics[width=\textwidth]{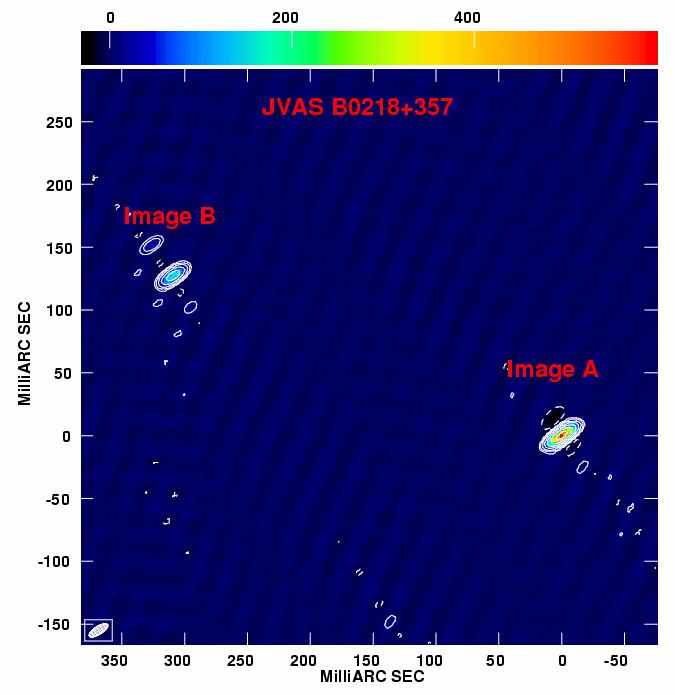}
\end{minipage}\hfill\raisebox{0.5ex}{\begin{minipage}{0.52\textwidth}
\ifpdf%
\includegraphics[angle=-90,width=\textwidth]{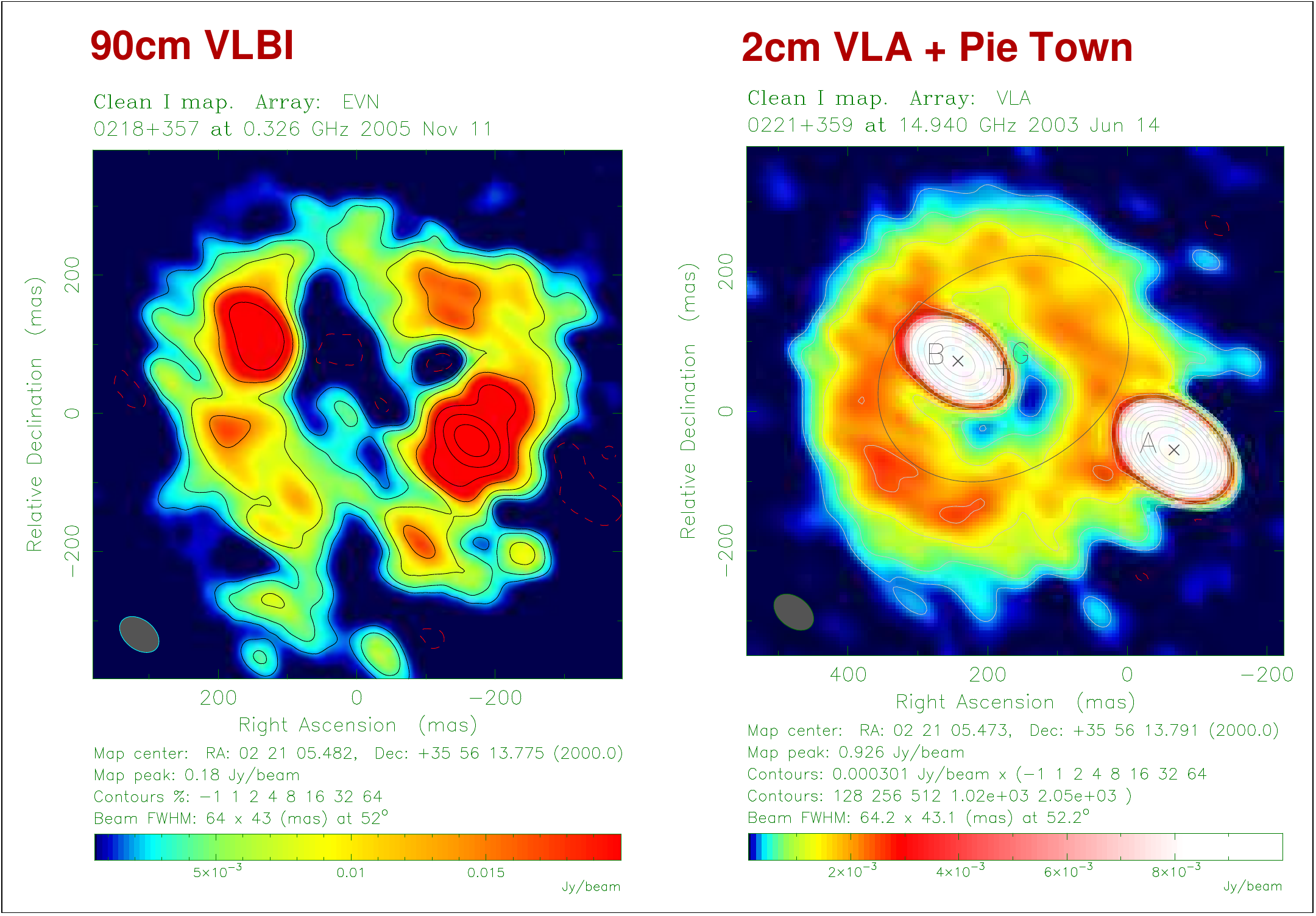}%
\else%
\includegraphics[width=\textwidth]{dailyimage}%
\fi%
\end{minipage}
}
\caption{Left: The first real-time EVN image produced at JIVE. Right:
  Preliminary 90-cm VLBI map in comparison with a 2-cm VLA+Pie~Town
  map. This is shown only for illustration, the calibration is still very poor
  at this stage.}
\label{fig:0218 more VLBI}
\end{figure}

It should be noted that B0218+357 was the target of the first real-time EVN
observations in 2004. The map from this small data set is shown in
Fig.~\ref{fig:0218 more VLBI} (left). Currently, the author is working on the
analysis of a 90-cm VLBI experiment conducted recently. The goal was to map
more details in the Einstein ring and to obtain more information about
propagation effects that are strongest at lower frequencies.
Fig.~\ref{fig:0218 more VLBI} (right) shows a very preliminary map compared to
a VLA+Pie~Town map at 2-cm.  Two additional correlator passes of the same
90-cm observations are used for a wide-field mapping experiment that
constitutes the first wide-field VLBI mini-survey to study the high-resolution
radio sky at such low frequencies. First results are presented by
\citet{lenc06} at this conference. See also \citet{lenc07}.

\section{Outlook}

\begin{figure}[htb]
\centering
\includegraphics[width=0.3\textwidth]{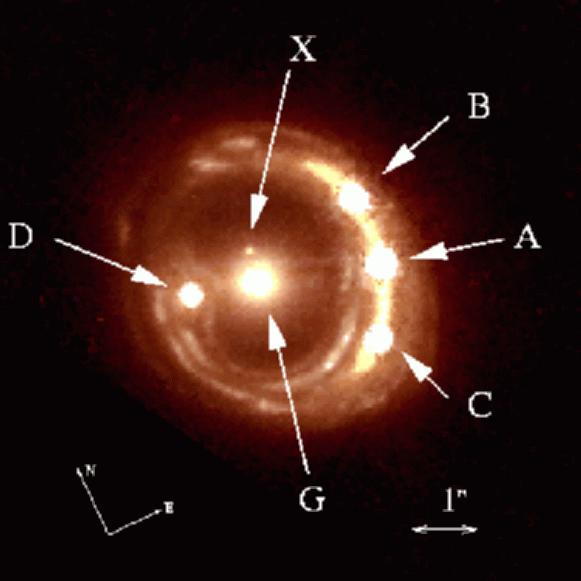}
\caption{Radio astronomy has to compete with this!
  Shown is the lensed star-burst galaxy J1131--1231 in the optical domain
  \citep[from][]{claeskens06}. Future radio telescopes will map such sources
  with even better quality than shown in this HST image. Each lensed
  star-forming region will provide its own constraints on the mass
  distribution of the lens.}
\label{fig:1131}
\end{figure}

Like many other fields, gravitational lens research takes advantage of the
technical progress in radio astronomy and particularly in VLBI. The EVN grows
continuously, providing steadily improving $uv$ coverage, mapping quality and
resolution. Higher bandwidths provide the increasing sensitivity that is
needed to map more extended source components with high quality. Such future
VLBI experiments will lead to more accurate lens mass models than currently
available.
This development is complemented by the upgraded EVLA and \emph{e}-MERLIN that
will come online soon. All these arrays will allow the mapping of even weaker
structures on scales of the image splitting with sufficient resolution to
utilise the information such sources provide for mass models and other
purposes. It will finally be possible to map even normal star-forming
galaxies in great detail. Such lensed sources have such a wealth of smaller
structures (Fig.~\ref{fig:1131}) that 
they allow a detailed mapping of the lensing potential and thus the mass
distribution of lensing galaxies.

EVLA, \emph{e}-MERLIN, and above all LOFAR (with extended baselines) will
allow future lens surveys that increase the number of radio lenses by at least
an order of magnitude.  This development has to be supplemented by the
development of new analysis techniques that can extract all information from
current and future observations. LensClean can only be one first step in this
direction.

\section{Summary}

We have discussed how the lens effect can be used to study all aspects of the
lensing situation. We use the lens as a natural telescope to study the sources
in greater detail, we determine the mass distributions of high-redshift lens
galaxies, and can even determine the Hubble constant and do cosmology.
Additionally, lenses provide us with (almost) identical copies of lensed
sources that can then be used to study differential propagation effects like
scattering or absorption.

Radio observations are particularly useful for several reasons. Firstly, radio
interferometers span the largest range of possible resolutions down to the
sub-mas level. Secondly, the effects of microlensing and extinction, that make
the interpretation of optical observations very difficult, can mostly be
avoided at radio wavelengths.  Instead of asking what radio astronomy can do
for lensing, we can also ask what lensing can do for radio astronomy. It is
our hope that the further development of deconvolution methods for the lensed
situation will also provide new and better algorithms for a general use. Such
a trigger for new developments should be very welcome.

Finally, we have to add that this review covers only a small subset of
relevant topics in gravitational lens research. Other important subjects like
microlensing, weak lensing, lensing surveys, measuring time delays, and a
number of very interesting lens systems have not been mentioned at all. That
does not imply that those fields are less exciting.

\section*{Acknowledgments}
This work was supported by the European Community's Sixth Framework
Marie Curie Research Training Network Programme, Contract No.\
MRTN-CT-2004-505183 ``ANGLES''.

\bibsep0.8ex
\bibliographystyle{proceedings}
\bibliography{proceedings}

\end{document}